\documentclass[aps,prl,reprint,superscriptaddress]{revtex4-2}
\usepackage{amsmath,amssymb,graphicx,bm}
\usepackage{xcolor}
\usepackage{ulem}
\begin{document}

\title{Probing the Spatial Coherence of Chiral-Imbalance Fluctuations with Hyperon Spin Correlations}

\author{Guangyu Zheng}
\email{zhengguangyu23@mails.ucas.ac.cn}
\affiliation{School of Nuclear Science and Technology, University of Chinese Academy of Sciences, Beijing 100049, P. R. China}
\author{Zilin Yuan}
\email{yuanzilin@ucas.ac.cn}
\affiliation{School of Nuclear Science and Technology, University of Chinese Academy of Sciences, Beijing 100049, P. R. China}
\author{Mei Huang}
\email{huangmei@ucas.ac.cn}
\affiliation{School of Nuclear Science and Technology, University of Chinese Academy of Sciences, Beijing 100049, P. R. China}
\author{Guannan Xie}
\email{xieguannan@ucas.ac.cn}
\affiliation{School of Nuclear Science and Technology, University of Chinese Academy of Sciences, Beijing 100049, P. R. China}

\begin{abstract}
Topological fluctuations in the quark-gluon plasma can generate local chiral imbalances whose spatial coherence remains difficult to access with conventional charge-separation measurements. We propose differential hyperon-pair spin correlations as a probe of both the strength and spatial organization of these fluctuations. 
Fluctuations of the axial chemical potential $\mu_5$ can induce pair correlations even when the mean helicity polarization vanishes.
Using the AMPT model, we find that the magnitude of the hyperon spin correlation probes the magnitude and spatial coherence scale of $\mu_5$, while the ratio of short- and long-range correlations varies strongly with the spatial coherence scale, but only weakly with $\mu_{5}$ over the range of explored parameters. These observables provide an experimentally testable strategy for constraining the spatial structure of chiral-imbalance fluctuations.

\end{abstract}

\maketitle

\textbf{Introduction.---}
Topological transitions of non-Abelian gauge fields can generate fluctuating axial charge through the chiral anomaly, giving rise to local chiral imbalance in the quark-gluon plasma (QGP)~\cite{Kharzeev:1998kz,Kharzeev:1999cz,Kharzeev:2001ev,Kharzeev:2004ey}. An effective axial chemical potential, $\mu_5$, characterizes this imbalance in a near-equilibrium description. In the strong magnetic fields generated in relativistic heavy-ion collisions, chiral imbalance can drive the chiral magnetic effect (CME)~\cite{Fukushima:2008xe,Kharzeev:2024zzm}. Experimental searches for the CME have extensively employed charge-dependent azimuthal correlations, particularly the $\Delta\gamma$ observable~\cite{Voloshin:2004vk,STAR:2021mii,STAR:2014uiw,ALICE:2012nhw}. However, substantial backgrounds from local charge conservation and resonance decays complicate the extraction of the CME signal~\cite{Bzdak:2009fc,Wang:2009kd,Schlichting:2010qia,Wu:2022fwz}.

Beyond constraining the overall strength of axial-charge fluctuations, an important open question is how chiral-imbalance fluctuations are correlated in space~\cite{Kharzeev:2024zzm,Voloshin:2004vk,Kharzeev:2001ev}. An event-wise uniform chiral imbalance and spatially fluctuating chiral domains can have the same fluctuation strength but distinctly different spatial correlations. Conventional CME-sensitive observables, which generally integrate over broad regions of the evolving collision system, do not directly resolve these spatial correlations. This motivates the search for complementary observables sensitive to the spatial organization of axial-charge fluctuations.

Hyperon-pair spin correlations offer a promising probe of this structure. Local chiral imbalance can bias quark helicity populations. Although sign-symmetric chiral fluctuations can yield vanishing event-averaged spin polarization, two-particle spin correlations can remain finite. If this partonic response is transferred to final-state $\Lambda$ and $\bar{\Lambda}$ spins, their parity-violating weak decays make the resulting spin correlations experimentally accessible through joint decay-angular distributions~\cite{STAR:2007ccu,Gong:2021bcp,STAR:2025njp}. Their dependence on the momentum-space pair separation $\Delta R=\sqrt{(\Delta\eta)^2+(\Delta\phi)^2}$ can retain information about the spatial organization of chiral fluctuations, because transport and collective expansion correlate final-state momenta with emission positions. However, $\Delta R$ is not a direct measure of the spatial separation between chiral domains.

In this Letter, we investigate differential hyperon-pair spin correlations using the AMPT model supplemented with chiral anomaly transport~\cite{Lin:2004en,Yuan:2023skl,Yuan:2024wpz}. In the studied kinematic region, the $\mu_5$-induced correlations substantially exceed the thermal-vorticity-plus-shear baseline, and the $\Lambda\bar{\Lambda}$ correlation closely follows the average of the $\Lambda\Lambda$ and $\bar{\Lambda}\bar{\Lambda}$ correlations. This motivates a common-mode observable combining all three pair channels. We further find that the ratio of short- to long-range correlations depends strongly on the effective spatial coherence scale but only weakly on the imposed $\mu_5$ magnitude within the parameter range explored. Together with the absolute correlation strength, this ratio provides complementary, experimentally testable constraints on the spatial organization of chiral-imbalance fluctuations.

\textbf{Physical mechanism and measurable spin correlations.---}

In a parity-symmetric event ensemble, configurations with
opposite topological charges have no preferred sign.
Consequently, the event-averaged chiral imbalance vanishes,
while its local fluctuations and spatial correlations
can remain finite.
In the massless-quark and weak-magnetic-field limits,
the axial-charge density $n_5=n_R-n_L$ is related to
$\mu_5$ by
 $n_5=\frac{\mu_5^3}{3\pi^2}
 +\frac{\mu_5}{3}
 \left(T^2+\frac{\mu^2}{\pi^2}\right),$
where $T$ and $\mu$ denote the temperature and quark
chemical potential, respectively, with the normalization
convention of the cited calculation~\cite{Fukushima:2008xe,Chao:2013qpa}.
This relation illustrates how a local $\mu_5$ generates
a local axial-charge imbalance.

For a quark or antiquark species $a$,
we define the local helicity asymmetry as
\begin{equation}
 H_a(\bm x,\bm p)=
 \frac{f_{a,+}(\bm x,\bm p)-f_{a,-}(\bm x,\bm p)}
 {f_{a,+}(\bm x,\bm p)+f_{a,-}(\bm x,\bm p)},
 \label{eq:helicity_bias}
\end{equation}
where $f_{a,\pm}$ are the phase-space distributions
for helicity $h=\pm1$.
In the massless limit, axial-charge imbalance is directly
related to the imbalance of helicity populations, with the
appropriate particle and antiparticle conventions.
For the finite-mass strange quarks considered here,
$H_a$ describes the helicity response to the local chiral
field and is not assumed to be linearly proportional
to $\mu_5$.

For a relativistic quark in a helicity eigenstate
$h=\bm\sigma\cdot\hat{\bm p}=\pm1$, the spin is parallel or antiparallel to its momentum,
$\bm s=(h/2)\hat{\bm p}$.
For two such particles,
\begin{equation}
 \bm s_a\cdot\bm s_b
 =\frac14 h_ah_b\cos\theta_{ab},
 \label{eq:spin_helicity_corr}
\end{equation}
where $\theta_{ab}$ is the three-dimensional opening
angle between their momenta.

For two spin-$1/2$ particles, we define the scalar
spin correlation as \cite{Lv:2024uev,STAR:2025njp},
\begin{equation}
 P_{ab} =\frac43\left\langle
 \bm s_a\cdot\bm s_b
 \right\rangle
 =\frac13\sum_{i=x,y,z}C_{ii}^{ab},
 \label{eq:pola_definition}
\end{equation}
where
$C_{ij}^{ab}=\langle\sigma_i^{(a)}\otimes\sigma_j^{(b)}
\rangle$ is the spin-correlation tensor, with consistently
defined spin axes.

For a fixed chiral-field configuration and fixed phase-space
coordinates, we assume that the helicities of the two
particles are independently sampled.
Their conditional expectation value then factorizes as
$\langle h_ah_b\rangle_{\mu_5}
 =\langle h_a\rangle_{\mu_5}
  \langle h_b\rangle_{\mu_5}
 =H_aH_b$, where subscript ${\mu_5}$ indicates finite chiral-field configuration.
Averaging over field configurations gives
\begin{equation}
 \langle h_ah_b\rangle
 =\left\langle
 H_a(\bm x_a,\bm p_a)
 H_b(\bm x_b,\bm p_b)
 \right\rangle_{\rm events}.
 \label{eq:helicity_factorization}
\end{equation}
Thus, conditional independence does not imply the absence
of an ensemble-averaged correlation: both particles can
respond to the same fluctuating chiral field.
The event-averaged helicity polarization can vanish for
sign-symmetric fluctuations, while $\langle H_aH_b\rangle$
remains finite.

\begin{figure}[t]
 \centering
 \includegraphics[width=\columnwidth]{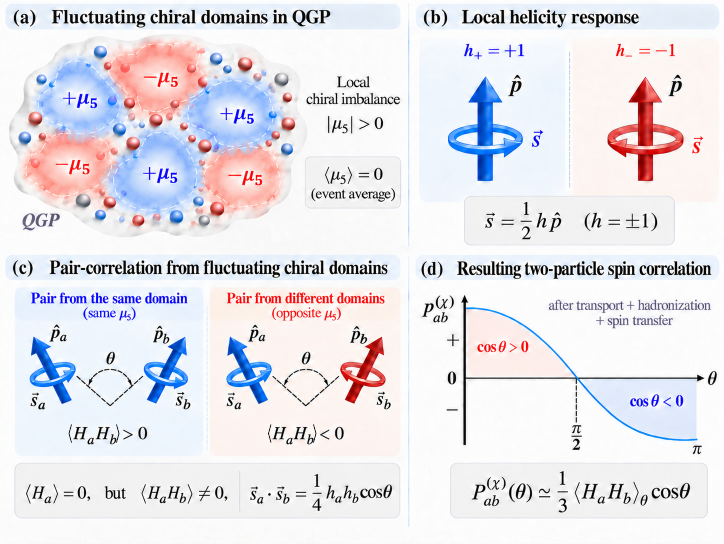}
 \caption{
 Schematic illustration of chiral-imbalance-induced
 spin correlations.
 (a) Spatial domains with opposite local signs of
 $\mu_5$ in the QGP, with $\langle\mu_5\rangle=0$
 over the event ensemble.
 (b) Helicity--spin locking: the spin is parallel
 (antiparallel) to the quark momentum for
 $h=+1$ ($h=-1$).
 (c) Same-sign and opposite-sign local helicity biases
 give positive and negative products $\langle H_aH_b\rangle$,
 respectively, for the illustrated configurations.
 For independent domains, the cross-domain correlation
 vanishes after event averaging.
 (d) Schematic angular response
 $P_{ab}^{(\chi)}(\theta)\simeq
 \langle H_aH_b\rangle_\theta\cos\theta/3$.
 For a positive helicity correlation, the spin-correlation
 sign changes with $\cos\theta$, rather than being fixed
 by the domain assignment.
 The final-state hyperon correlation requires a dynamical
 mapping from the partonic response.
 }
 \label{fig:mechanism2}
\end{figure}

Using Eq.~(\ref{eq:spin_helicity_corr}), the contribution
induced by chiral-imbalance fluctuations, denoted by the
superscript $(\chi)$, takes the schematic form
\begin{equation}
 P_{ab}^{(\chi)}(\theta_{ab})
 \simeq \frac13
 \left\langle h_ah_b\cos\theta_{ab}
 \right\rangle_{\theta_{ab}}
 \simeq \frac13
 \left\langle H_aH_b\right\rangle_{\theta_{ab}}
 \cos\theta_{ab},
 \label{eq:schematic_pair}
\end{equation}
where subscript $\theta_{ab}$ indicates the averages are taken over events and particle pairs
at a specified opening angle $\theta_{ab}$.
The angular factor reflects helicity--spin locking, while
$\langle H_aH_b\rangle_{\theta_{ab}}$ encodes the correlated
helicity response to the fluctuating chiral field.
This mechanism does not require direct pairwise spin
interactions or quantum entanglement between the particles.
The physical picture is illustrated in Fig.~\ref{fig:mechanism2}.

To connect the partonic mechanism to experimentally
accessible hyperons, we adopt the conventional SU(6)
constituent-quark picture, in which the $ud$ pair in a
directly produced $\Lambda$ carries predominantly zero
spin and the hyperon spin is carried by its constituent
$s$ quark; the corresponding assumption applies to
$\bar\Lambda$ and $\bar s$~\cite{Liang:2004ph,Yang:2017sdk}.
Accordingly, the spin response of a directly produced
$\Lambda$ ($\bar\Lambda$) is taken to follow that of its
constituent $s$ ($\bar s$) quark immediately before
coalescence, with the spin vectors expressed in a
consistent basis.
This is a spin-transfer prescription in the numerical
realization, rather than an exact identity for all
hyperons; subsequent hadronic evolution can modify
the resulting correlations.

We study the final-state hyperon-pair correlation
$P_{AB}(\Delta R)$, where
$A,B\in\{\Lambda,\bar\Lambda\}$ and
 $\Delta R=\sqrt{(\Delta\eta)^2+(\Delta\phi)^2},
-\pi\leq\Delta\phi\leq\pi.$
Here, $\eta$ is pseudorapidity and $\Delta\phi$ is the
wrapped relative azimuthal angle.
The dependence on $\Delta R$ is motivated by two
complementary considerations.

First, $\Delta R$ is kinematically related to the opening
angle $\Theta_{AB}$ between the final-state hyperon momenta.
For small angular separations,
$\Theta_{AB}\simeq\Delta R/\cosh\bar\eta$, where
$\bar\eta=(\eta_A+\eta_B)/2$.
Consequently, differential measurements in $\Delta R$
retain information about the angular dependence in
Eq.~(\ref{eq:schematic_pair}), although the final-state
hyperon opening angle is not identical to the partonic
angle $\theta_{ab}$.

Second, transport and collective expansion establish
statistical correlations between emission positions and
final-state momenta~\cite{Sorensen:2010zq}.
Pairs selected in different $\Delta R$ intervals can
therefore sample different distributions of source
separations and, consequently, different correlations
of the chiral field.
If larger $\Delta R$ preferentially selects particles
from more widely separated source regions, finite spatial
coherence can suppress their helicity-pair response
relative to that of small-$\Delta R$ pairs.
This space--momentum mapping is dynamical and model
dependent: $\Delta R$ is not a direct measure of the
coordinate-space separation between chiral domains.
The interplay of angular response and spatial coherence
motivates studying differential rather than only
pair-integrated spin correlations.

Experimentally, the scalar hyperon-pair correlation
can be extracted from the acceptance-corrected joint
weak-decay angular distribution,
 $\frac1N\frac{dN}{d\cos\theta^*}
 =\frac12\left[
 1+\alpha_A\alpha_B
 P_{AB}(\Delta R)\cos\theta^*
 \right],$
where $\alpha_A$ and $\alpha_B$ are the signed weak-decay
parameters, and $\theta^*$ is the angle between the
daughter (anti)proton momentum directions in their
respective hyperon rest frames~\cite{STAR:2025njp}.
The measured $P_{AB}$ is the total spin correlation from
all sources; isolating the contribution from chiral
fluctuations requires controlling other sources of
pair-spin correlations.

\textbf{Experimentally benchmarked numerical realization.---}
To study the proposed spin-correlation response, we employ
the string-melting AMPT model~\cite{Lin:2004en} supplemented
with the Chiral Anomaly Transport (CAT) module
~\cite{Yuan:2023skl,Yuan:2024wpz} for Au+Au collisions at
$\sqrt{s_{NN}}=200$ GeV.
AMPT describes partonic evolution, quark coalescence, and
hadronic rescattering, while CAT incorporates chiral transport
in the presence of electromagnetic fields and axial-charge
imbalance~\cite{Stephanov:2012ki, Huang:2018wdl,Zhou:2018rkh}.
Following the spin-transfer prescription described above,
we assign each $\Lambda$ ($\bar\Lambda$) the polarization of
its constituent $s$ ($\bar s$) quark during quark coalescence~\cite{Liang:2004ph}.
Thermal-vorticity and shear ~\cite{Fu:2021pok,Wu:2022mkr,Yi:2021ryh} contributions are calculated
separately from the local flow and temperature fields,
providing a conventional spin-polarization baseline.

Rather than dynamically generating individual topological
transitions, we impose controlled $\mu_5$ configurations to
study how their spatial correlations are reflected in
final-state hyperon spins.
We consider two limiting cases: an event-wise uniform field,
$\mu_5(\bm x)=s\,\mu_{5,0}$, where $s=\pm1$ is sampled once per event; and a spatially fluctuating
domain field,
 $\mu_5(\bm x_i)=s_i\,\mu_{5,0}$,where $s_i=\pm1$ is independently sampled for each spatial cell, both with equal probabilities.The two configurations share the same ensemble mean and
local variance,
$\langle\mu_5\rangle=0$ and
$\langle\mu_5^2\rangle=\mu_{5,0}^2$,
but differ in their spatial correlations.

We characterize the spatial correlations of the imposed field through
the two-point covariance
$C_5(\bm x_a,\bm x_b)=
\langle\mu_5(\bm x_a)\mu_5(\bm x_b)\rangle$,
where the average is taken over field configurations and
$\langle\mu_5(\bm x)\rangle=0$.
Here, $\bm x_a$ and $\bm x_b$ denote the locations at which
the two particles sample the chiral field.
For the event-wise uniform field,
$C_5(\bm x_a,\bm x_b)=\mu_{5,0}^2$ throughout the system.
For finite-size domains with independently sampled signs,
$C_5=\mu_{5,0}^2$ within the same cell and vanishes between
different cells.
Although the field product between different cells is
$\pm\mu_{5,0}^2$ in an individual event, its ensemble average is zero.
The cell size $l_{\rm cell}$ therefore controls the effective
spatial coherence of the imposed field, while the uniform
configuration represents the long-coherence limit.
Here, $l_{\rm cell}$ is a phenomenological coarse-graining scale
rather than a direct measure of the microscopic sphaleron size.

\begin{figure}[t]
 \centering
 \includegraphics[width=\columnwidth]{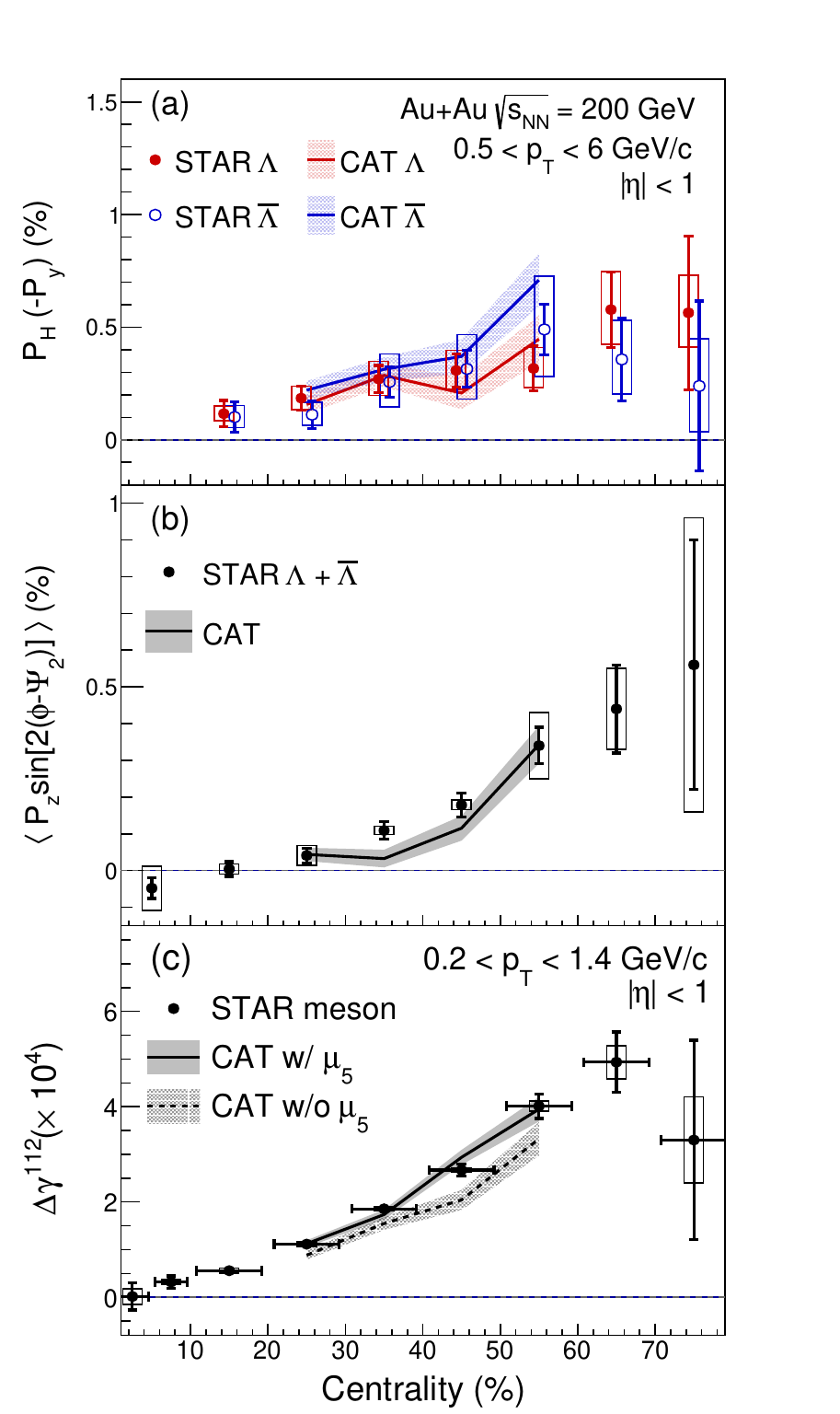}
 \caption{
 Comparison of the numerical calculations with STAR measurements
 in Au+Au collisions at $\sqrt{s_{NN}}=200$ GeV.
 (a) Global polarization of $\Lambda$ and $\bar\Lambda$ hyperons\cite{STAR:2018gyt}.
 (b) Longitudinal polarization modulation of $\Lambda+\bar\Lambda$\cite{STAR:2019erd}.
 These observables benchmark the conventional spin response.
 (c) CME-sensitive charge-dependent correlator
 $\Delta\gamma^{112}$\cite{STAR:2025vhs}, compared with calculations with and
 without finite $\mu_5$.
 }
 \label{fig:constraints}
\end{figure}

Figure~\ref{fig:constraints} benchmarks complementary
ingredients of the numerical framework against existing
STAR measurements.
The global and longitudinal polarization observables~\cite{STAR:2017ckg,Becattini:2016gvu,STAR:2019erd,Becattini:2020ngo}
constrain the conventional spin response, including
contributions from thermal vorticity and shear~\cite{Huang:2020dtn,Wei:2018zfb}.
For the CME-sensitive $\Delta\gamma^{112}$ observable,
including finite $\mu_5$ improves the description of the
measured correlation in the present calculation.
However, the substantial conventional backgrounds
contributing to $\Delta\gamma^{112}$ prevent this comparison
from uniquely determining the axial-charge fluctuation
strength.
Importantly, these observables do not independently resolve
the spatial correlations of the axial-charge field.
Differential hyperon-pair spin correlations can therefore
provide complementary information on the spatial structure
of chiral-imbalance fluctuations.

\begin{figure}[t]
 \centering
 \includegraphics[width=\columnwidth]{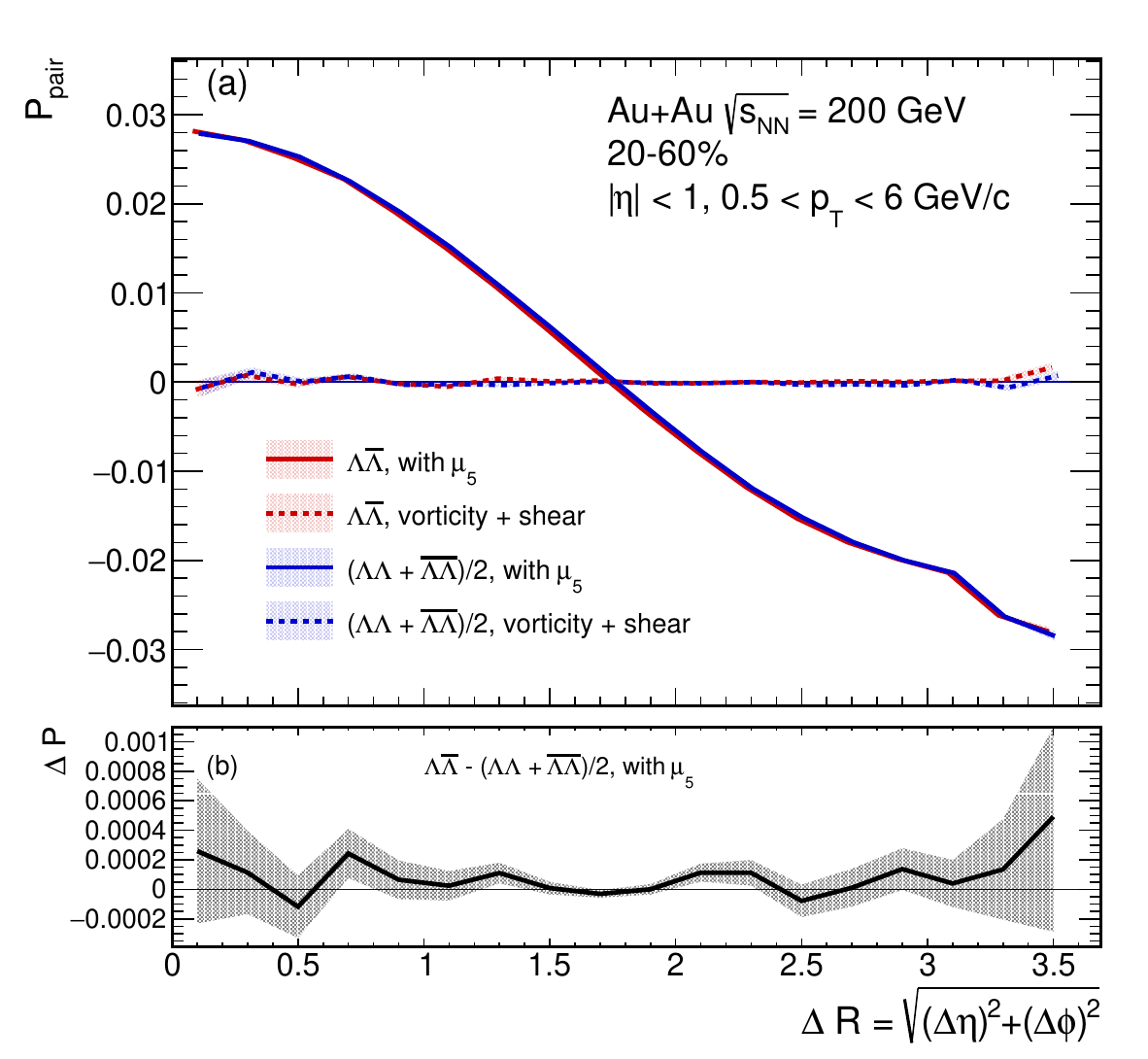}
 \caption{
 Differential hyperon-pair spin correlations in
 $20$--$60\%$ Au+Au collisions at
 $\sqrt{s_{NN}}=200$ GeV, for $|\eta|<1$ and
 $0.5<p_T<6$ GeV/$c$.
 (a) $P_{\Lambda\bar{\Lambda}}$ and the average of
 $P_{\Lambda\Lambda}$ and $P_{\bar{\Lambda}\bar{\Lambda}}$,
 with finite $\mu_5$ (solid lines) and for the
 thermal-vorticity-plus-shear baseline (dashed lines).
 (b) Difference $\Delta P$ defined in Eq.~(\ref{eq:deltaP}).
 }
 \label{fig:channels}
\end{figure}

\textbf{A common-mode response across pair channels.---}
Figure~\ref{fig:channels}(a) shows the differential hyperon-pair
spin correlations in $20$--$60\%$ Au+Au collisions at
$\sqrt{s_{NN}}=200$ GeV, for $|\eta|<1$ and
$0.5<p_T<6$ GeV/$c$.
At $\mu_5=0$, the baseline spin correlations generated by thermal
vorticity and shear are substantially smaller than the finite-$\mu_5$
results over most of the investigated range, consistent with previous
findings~\cite{Pang:2016igs}.
We therefore neglect the thermal-vorticity and shear contributions
in the following analysis and focus on the correlations induced by
chiral-imbalance fluctuations. With finite $\mu_5$, the correlations  change from positive at small $\Delta R$ to negative at large $\Delta R$.
This demonstrates the strong sensitivity of the pair observable
to chiral imbalance within the present transport framework.

An additional feature is the approximate channel independence
of this response: $P_{\Lambda\bar{\Lambda}}$ closely follows
the average of $P_{\Lambda\Lambda}$ and
$P_{\bar{\Lambda}\bar{\Lambda}}$.
We quantify this consistency through
\begin{equation}
 \Delta P(\Delta R)=
 P_{\Lambda\bar{\Lambda}}(\Delta R)
 -\frac{
 P_{\Lambda\Lambda}(\Delta R)+
 P_{\bar{\Lambda}\bar{\Lambda}}(\Delta R)
 }{2},
 \label{eq:deltaP}
\end{equation}
which is consistent with zero within statistical uncertainties
[Fig.~\ref{fig:channels}(b)].
This agreement motivates combining the three channels into
a common-mode observable.

We therefore define the common-mode spin correlation as
\begin{equation}
 P_{\mathrm{common}}(\Delta R)=\frac14
 \left[
 P_{\Lambda\Lambda}(\Delta R)
 +2P_{\Lambda\bar{\Lambda}}(\Delta R)
 +P_{\bar{\Lambda}\bar{\Lambda}}(\Delta R)
 \right].
 \label{eq:pcommon}
\end{equation}
The fixed $1:2:1$ combination preserves a channel-independent
signal, $P_{ab}^{(\chi)}(\Delta R)\simeq S(\Delta R)$,
while canceling a background with the specific sign structure.
$P_{ab}^{(B)}(\Delta R)=q_aq_b B(\Delta R)$, where
$q_\Lambda=+1$ and $q_{\bar{\Lambda}}=-1$.
For this background, $P_{\mathrm{common}}^{(B)}=0$,
whereas $\Delta P^{(B)}=-2B$.
The combination thus retains the common chiral response
while providing selective suppression of particle--antiparticle
sign-dependent backgrounds. Such as spin correlation induced by the strong-force fields\cite{Sheng:2025puj}.
It does not, however, remove backgrounds shared by all three
channels, including possible hadronization-induced correlations.
The coefficients define a fixed linear combination of the
channel-level correlations, rather than a weighting by the
observed pair yields.

\textbf{Complementary sensitivity to fluctuation strength and spatial coherence.---}
Having established the common response across hyperon-pair
channels, we investigate how its magnitude and differential
structure encode the properties of chiral-imbalance fluctuations.
We select two fixed, nonoverlapping intervals,
$0\leq\Delta R<1.0$ and $2.5\leq\Delta R<3.5$,
to characterize short- and long-range correlations.
For a given pair channel $ab$ and interval $\Omega$, we define
the pair-weighted correlation as
\begin{equation}
 P_{ab}^{\Omega}=
 \frac{\int_{\Omega}d(\Delta R)\,
 \frac{dN_{ab}}{d\Delta R}P_{ab}(\Delta R)}
 {\int_{\Omega}d(\Delta R)\,
 \frac{dN_{ab}}{d\Delta R}},
 \label{eq:integrated_pair}
\end{equation}
where $dN_{ab}/d\Delta R$ is the differential pair yield.
The common-mode correlation in each interval is obtained by
applying Eq.~(\ref{eq:pcommon}) to the three channel-averaged
correlations. We denote the resulting observables by
 $P_{\rm short}=P_{\rm common}^{\,0\leq\Delta R<1.0},
 P_{\rm long}=P_{\rm common}^{\,2.5\leq\Delta R<3.5},$
and define their ratio as
\begin{equation}
 \mathcal R_{SL}=-\frac{P_{\rm short}}{P_{\rm long}}.
 \label{eq:ratio_def}
\end{equation}

The angular factor in Eq.~(\ref{eq:schematic_pair}) changes
sign at $\theta_{ab}=\pi/2$ when the helicity-pair response
remains finite.
However, the partonic opening angle $\theta_{ab}$ does not
map uniquely onto the final-state hyperon separation $\Delta R$,
owing to the momentum distributions, transport, and hadronization.
We therefore use fixed intervals that sample the positive and
negative regions of the calculated correlation, avoiding the
need to define a model-dependent zero-crossing point.
These intervals provide a consistent basis for comparing
different configurations of the chiral field.

\begin{figure}[t]
 \centering
 \includegraphics[width=\columnwidth]{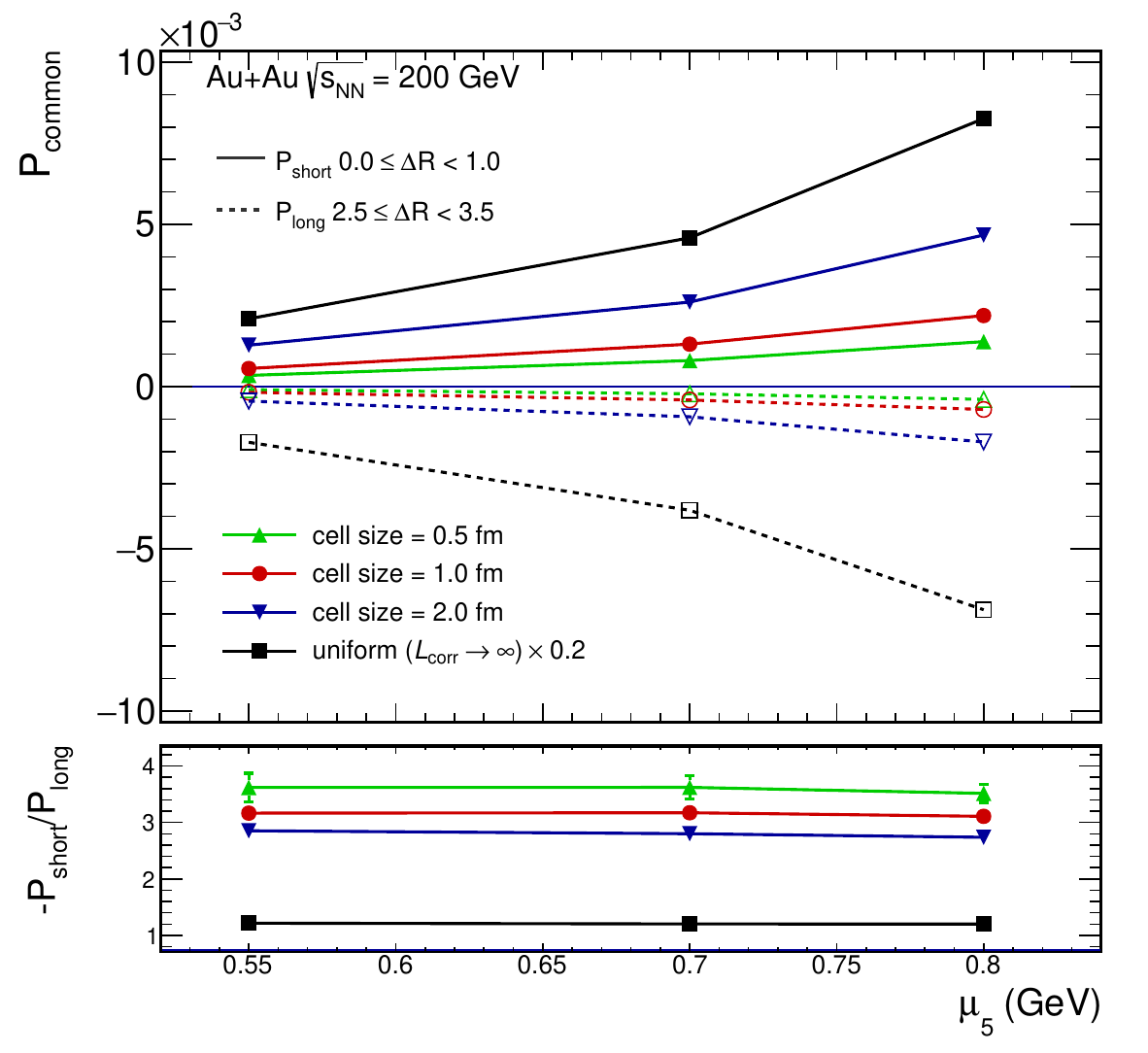}
 \caption{
 Dependence of the common-mode spin correlation on the magnitude
 and spatial coherence of chiral-imbalance fluctuations in
 Au+Au collisions at $\sqrt{s_{NN}}=200$ GeV.
 (a) Short-range ($0\leq\Delta R<1.0$, solid lines) and
 long-range ($2.5\leq\Delta R<3.5$, dashed lines) correlations
 as functions of $\mu_{5,0}$ for cell sizes of $0.5$, $1.0$,
 and $2.0$ fm and the event-wise uniform limit.
 The uniform results in panel (a) are scaled by $0.2$ for
 display only.
 (b) The unscaled short-to-long-range ratio
 $\mathcal R_{SL}=-P_{\rm short}/P_{\rm long}$.
 The ratio exhibits a pronounced dependence on the effective
 spatial coherence scale but only weak variation with
 $\mu_{5,0}$ over the investigated range.
 }
 \label{fig:coherence}
\end{figure}

Figure~\ref{fig:coherence}(a) shows the dependence of
$P_{\rm short}$ and $P_{\rm long}$ on the imposed magnitude
$\mu_{5,0}$ for different spatial coherence scales.
The short-range correlation is positive, whereas the long-range
correlation is negative, consistent with the angular dependence
of the proposed mechanism.
For finite domains, both correlations increase in absolute
magnitude with $\mu_{5,0}$ and generally with the cell size
$l_{\rm cell}$.
The event-wise uniform configuration produces substantially
larger correlations, illustrating the dilution of the net
spin response by spatially fluctuating domain signs.
For visibility, the uniform results in panel (a) are multiplied
by $0.2$; this factor is not applied to the ratio.

The ratio in Fig.~\ref{fig:coherence}(b) exhibits a different
sensitivity.
Over the investigated range
$\mu_{5,0}=0.55$--$0.80$ GeV, $\mathcal R_{SL}$ depends only
weakly on the imposed fluctuation magnitude but shows a clear
dependence on the spatial coherence scale.
It decreases from approximately $3.5$--$3.6$ for
$l_{\rm cell}=0.5$ fm to about $2.8$ for $2.0$ fm,
while the uniform configuration yields approximately $1.2$.
Thus, more localized chiral fluctuations suppress the
long-range anticorrelation more strongly relative to the
short-range positive correlation.
The pronounced contrast between the uniform and finite-domain
configurations further demonstrates the sensitivity of
$\mathcal R_{SL}$ to the spatial organization of the chiral field.

The weak dependence of $\mathcal R_{SL}$ on $\mu_{5,0}$
suggests an approximate separation between the fluctuation
amplitude and spatial response.
Phenomenologically, the numerical trends can be expressed as
\begin{equation}
 \begin{aligned}
 P_{\rm short}&\simeq
 A(\mu_{5,0})K_{\rm short}(l_{\rm cell}),\\
 P_{\rm long}&\simeq
 -A(\mu_{5,0})K_{\rm long}(l_{\rm cell}),
 \end{aligned}
 \label{eq:factorization}
\end{equation}
where $A$ characterizes the overall amplitude response and
$K_{\rm short,long}$ describe the effective spatial response
after transport and hadronization.
Consequently,
\begin{equation}
 \mathcal R_{SL}\simeq
 \frac{K_{\rm short}(l_{\rm cell})}
 {K_{\rm long}(l_{\rm cell})},
 \label{eq:ratio_factorization}
\end{equation}
with much of the common amplitude dependence canceling
in the ratio.
This factorization is a phenomenological interpretation of
the observed trends, not a universal relation.

These results suggest a complementary strategy for constraining
chiral-imbalance fluctuations: $\mathcal R_{SL}$ provides a
relatively amplitude-insensitive probe of the effective spatial
coherence, while the absolute correlation magnitudes retain
sensitivity to the fluctuation strength.
Their combined use can therefore help constrain both properties
within a specified dynamical framework.
Since $\mathcal R_{SL}$ is constructed from $P_{\rm short}$
and $P_{\rm long}$, a quantitative extraction must account
for the statistical and systematic covariance between them.

\textbf{Summary.---}
We propose differential hyperon-pair spin correlations as a probe
of the strength and spatial coherence of chiral-imbalance
fluctuations in the QGP.
In the present transport framework, chiral-imbalance fluctuations
generate a finite two-particle spin correlation even when the
event-averaged helicity polarization vanishes.
The induced correlation exceeds the thermal-vorticity and shear
contributions in the investigated kinematic range and exhibits
an approximately common response across the
$\Lambda\Lambda$, $\Lambda\bar\Lambda$, and
$\bar\Lambda\bar\Lambda$ channels.
This motivates the common-mode observable $P_{\rm common}$
and the channel-difference observable $\Delta P$.
We further find that the short-to-long-range ratio
$\mathcal R_{SL}=-P_{\rm short}/P_{\rm long}$ is strongly
sensitive to the effective spatial coherence scale but only
weakly dependent on the imposed $\mu_5$ magnitude over the
explored parameter range, while the absolute correlations
retain sensitivity to both.
These observables can be extracted from joint weak-decay angular
distributions and provide complementary, experimentally testable
constraints on chiral-imbalance fluctuations.
Their interpretation will require controlling other sources
of pair-spin correlations, including those associated with
hadronization and possible quantum correlations.
The inferred spatial coherence scale is model dependent and
should not be identified directly with the microscopic
sphaleron size.

\textbf{Acknowledgments.---}
This work is supported in part by the National Natural Science Foundation of China (NSFC) Grant Nos. 12235016, 12221005, 12305146 and by the CAS Project for Young Scientists in Basic Research under Grant No. YSBR-088.
% ...

\bibliographystyle{apsrev4-2}
\bibliography{reference}

\end{document}